\documentclass[10pt]{iopart}

\usepackage{iopams}
\usepackage{graphicx}
\begin{document}
\title{Inhomogeneous hysteresis in local STM tunnel conductance with gate-voltage in single-layer MoS$_2$ on SiO$_2$}
\author{Santu Prasad Jana, Suraina Gupta \& Anjan Kumar Gupta}
\address{Department of Physics, Indian Institute of Technology Kanpur, Kanpur 208016, India}
\ead{anjankg@iitk.ac.in}
\vspace{10pt}
\begin{indented}
\item[]{\today}
\end{indented}
\begin{abstract}
Randomly distributed traps at the MoS$_2$/SiO$_2$ interface result in non-ideal transport behavior, including hysteresis in MoS$_2$/SiO$_2$ field effect transistors (FETs). Thus traps are mostly detrimental to the FET performance but they also offer some application potential. Our STM/S measurements on atomically resolved few-layer and single-layer MoS$_2$ on SiO$_2$ show n-doped behavior with the expected band gap close to 2.0 and 1.4 eV, respectively. The local tunnel conductance with gate-voltage $V_{\rm g}$ sweep exhibits a turn-on/off at a threshold $V_{\rm g}$ at which the tip's Fermi-energy nearly coincides with the local conduction band minimum. This threshold value is found to depend on $V_{\rm g}$ sweep direction amounting to local hysteresis. The hysteresis is, expectedly, found to depend on both the extent and rate of $V_{\rm g}$-sweep. Further, the spatial variation in the local $V_{\rm g}$ threshold and the details of tunnel conductance Vs $V_{\rm g}$ behavior indicate inhomogenieties in both the traps' density and their energy distribution. The latter even leads to the pinning of the local Fermi energy in some regions. Further, some rare locations exhibit a p-doping with both p and n-type $V_{\rm g}$-thresholds in local conductance and an unusual hysteresis.
\end{abstract}
%
\vspace{2pc}

\submitto{\TDM}
%
\maketitle
%

\ioptwocol
\section{Introduction}
In recent years, transition metal dichalcogenides (TMDs) have gained significant interest for next-generation semiconductor devices \cite{TMDs}. Molybdenum disulfide (MoS$_2$) is a promising TMD with a direct bandgap of $\sim1.9$ eV in monolayer form and an indirect bandgap of $\sim1.3$ eV in bulk form \cite{direct gap,direct gap1}, making it an excellent candidate for transistor \cite{direct gap1,how good}, logic \cite{logic}, high-frequency \cite{high frequency}, circuit integration \cite{ic} and optoelectronic \cite{photodetectors,light,photo,direct gap1} applications. However, MoS$_2$ single layer exhibits decreased charge carrier mobility compared to graphene due to its poor dielectric screening, inherent structural defects, and quantum confinement effects \cite{defect,defect1}. Interface traps between atomically thin MoS$_2$ and dielectric substrate and intrinsic defects are prevalent issues that impact electrical transport characteristics. For instance, a persistent photoconductivity is observed in MoS$_2$ photo-transistors \cite{photoconductivity} due to the photo-charge trapping by intrinsic localized band-tail states and extrinsic interface trap states. 

\par The traps are also responsible for the observed hysteresis in the transfer characteristics, i.e. drain-source current $I_{\rm DS}$ Vs gate voltage $V_{\rm g}$, of MoS$_2$ field effect transistors (FETs). This has been studied by several research groups \cite{acs nano,scalling behevior,hysteresis inversion,intrinsic origin,interface,cvd1,oxide traps,oxide traps close to Si,mobile ions,trap-block-trans} and as a function of gate bias stress, gate sweeping range, sweeping time, sweeping direction and loading history in different conditions such as high and low temperature, vacuum and different gas environments. Hysteresis reported in single and multi-layer MoS$_2$ has been ascribed to several possibilities including the absorption of water and gas molecules on top of MoS$_2$ \cite{acs nano,scalling behevior}, charge traps at the metal-semiconductor interface of the contacts \cite{hysteresis inversion}, the intrinsic structural defects \cite{intrinsic origin}, the extrinsic traps at the MoS$_2$/SiO$_2$ interface \cite{interface,cvd1}, oxide traps close to MoS$_2$/SiO$_2$ interface \cite{oxide traps}, oxide traps close to p$ ^{\rm +} $ Si/SiO$_2$ interface \cite{oxide traps close to Si} and mobile ions (Na$ ^{\rm +} $ and K$ ^{\rm +} $) in the oxide \cite{mobile ions}. In bulk transport, the threshold $V_{\rm g}$ value at which the channel conduction starts/stops and its change between the two opposite $V_{\rm g}$-sweep directions give an averaged information about the traps that are actually non-uniformly distributed. Traps also have potential in memory devices as some recent work illustrates \cite{trap-memory}.

\par In the absence of traps, the oxide capacitance and the MoS$_2$ quantum capacitance dictate the carrier density of the MoS$_2$ channel \cite{quantum capacitance} in response to $V_{\rm g}$. The presence of the interface charged-traps, lead to a spatially inhomogeneous electrostatic channel potential that dictates the local charge density and overall carrier mobility. A non-uniform density and energy-distribution of trap states lead to a spatially inhomogeneous, and $V_{\rm g}$ dependent, screening of the gate electric field. This can cause a non-uniform change in carrier density and unpredictable changes in mobility with $V_{\rm g}$. For instance, a local peak in traps' density of states can pin the MoS$_2$ Fermi energy keeping the carrier density unchanged over significant $V_{\rm g}$ range.

\par  In this paper, we report on hysteresis and its inhomogeneities with gate sweep in the local tunnel conductance of atomically resolved single and few-layer MoS$_2$ on SiO$_2$ by using a scanning tunneling microscope (STM). The measured tunnel conductance spectra, i.e. $dI/dV_{\rm b}$ Vs bias voltage $V_{\rm b}$, show the expected band-gap while the $dI/dV_{\rm b}$ Vs $V_{\rm g}$ curves, at a fixed $V_{\rm b}$, exhibit spatial inhomogeneities including that in $V_{\rm g}$-threshold. This threshold is found to differ during forward and backward sweeps of $V_{\rm g}$ with this difference being dependent on the extent and rate of $V_{\rm g}$-sweep. A rare p-doped region, confirmed from spectra, is found to show unusual hysteresis. These findings are discussed using traps with an inhomogeneous density and energy distribution.

\section{Experimental Details}
\subsection{MoS$_2$ Exfoliation, characterization and device fabrication }
Single and few-layer MoS$_2$ flakes were mechanically exfoliated from natural bulk crystal (from SPI) using a Scotch tape \cite{scotchtape method} and transferred on SiO$_2$/Si substrate by dry transfer technique \cite{XYZ} employing a PDMS membrane (Gel film from Gel Pak) as a viscoelastic stamp under an optical microscope. Acetone/IPA cleaned, highly doped Si acting as back gate with 300 nm thermal SiO$ _{2}$ is used as substrate. The number of MoS$_2$ layers on SiO$_2$/Si substrate was determined by optical microscope and Raman spectroscopy.
\begin{figure*}
	\centering
 	\includegraphics[width=16.6cm]{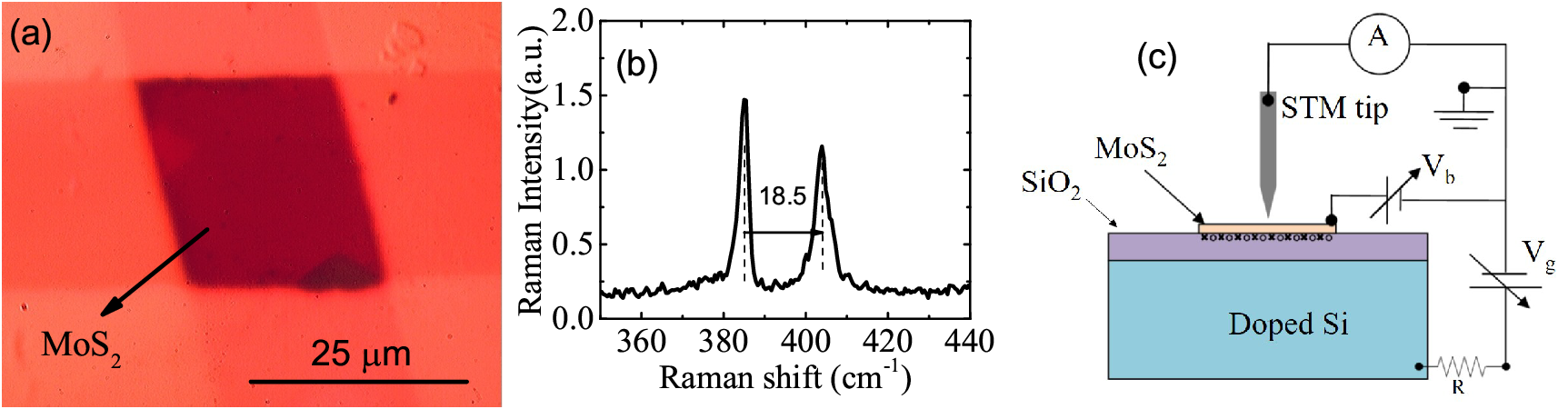}
	\caption{(a) Optical image of single-layer MoS$_2$ on SiO$_2$/Si substrate surrounded by gold contacts. The inset shows the Raman spectrum of an exfoliated single layer MoS$_2$ flake on SiO$_2$. (b) The electrical schematic of the STM measurement technique. Here, $V_{\rm b}$ is the bias voltage applied on sample keeping the tip at virtual ground and $V_{\rm g}$ is the back-gate voltage. Note that in this configuration and for $V_{\rm b}>0$ electrons tunnel from the tip's filled-states to the MoS$_2$ vacant-states.}
	\label{fig:mos21}
\end{figure*}
Figure \ref{fig:mos21}(b) shows a separation between the  E$^{\rm 1}_{\rm 2g} $ and A$ _{\rm 1g} $ Raman peaks as 18.5 cm$ ^{\rm -1}$, which is comparable to that reported \cite{optical identification,cvd} in single-layer MoS$_2$.

A clean surface, free of resist and wet chemicals' residue, is required for STM/S studies. Therefore, we use mechanical masking to make 50 nm thick gold film contacts by placing and aligning the MoS$_2$ flake underneath a 25 $\mu$m diameter tungsten wire using an optical microscope and then depositing Au. This is followed by a second wire alignment, nearly perpendicular to the previous, and Au deposition to surround MoS$_2$ with Au from all sides as shown in figure \ref{fig:mos21}(a). This helps in aligning the STM tip on MoS$_2$ under an optical microscope. The contact resistance was estimated to be about 5 $k\Omega$-$\mu$m$^2$ from separate multi-probe transport measurements on samples made by e-beam lithography.

\subsection{STM measurement details}
The STM/S measurements were done in a homemade room temperature STM in a cryo-pumped vacuum better than 10$^{\rm -4}$ mbar. Electrochemically etched and HF cleaned tungsten wire was used as STM tip with an apex radius between 20 and 50 nm as confirmed by scanning electron microscope. Figure \ref{fig:mos26}(b) depicts the STM/S measurement schematic with the bias voltage $V_{\rm b}$ applied to MoS$_2$ while the tip stays at virtual ground potential. Thus, at a positive sample bias electrons tunnel from the tip's filled-states to the vacant-states of MoS$_2$. The gate voltage $V_{\rm g}$ is applied to the heavily doped Si-substrate with a 10 k$\Omega $ series resistance. Before doing the STM/S measurements on MoS$_2$, we have taken atomic resolution images on HOPG sample to ensure good tip conditions.

To acquire the tunnel conductance $dI/dV_{\rm b}$ spectra, an AC modulation voltage of 20 mV and 2731 Hz frequency was added to the DC bias voltage and the in-phase component of the AC tunnel current was measured using a Lock-in amplifier. The $dI/dV_{\rm b}$ curves presented here are averages of twelve such curves unless stated otherwise. For the local $dI/dV_{\rm b}$ measurements, the feedback loop is turned off to keep a fixed tip-sample separation. A standby gate voltage $V_{\rm g}=30$ V is also used to ensure sufficient bulk conductivity of MoS$_2$ after the spectra acquisition is stopped and before the feedback loop is turned on. This used positive $V_{\rm g}$ is above the typical threshold voltage for bulk conduction as found from separate transport measurements.

\begin{figure*}
	\centering
	\includegraphics[width=15.6cm]{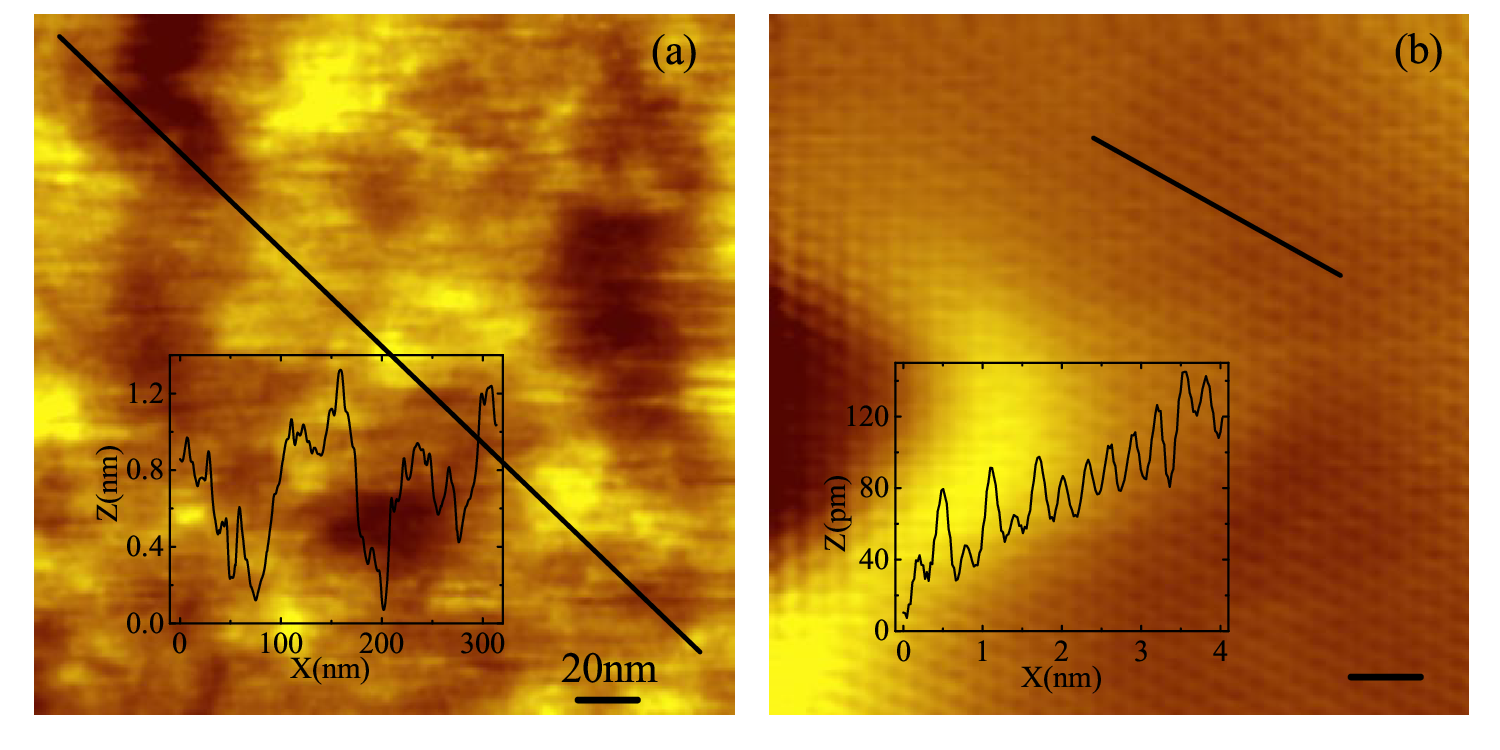}
	\caption{(a) STM topographic image taken at $V_{\rm b}=2$ V and 50 pA tunnel current of single layer MoS$_2$ on SiO$_2$/Si substrate at a gate voltage V$_{\rm g}=$ 30 V. The plot in the inset depicts the topography along the marked line showing the single layer MoS$_2$ corrugation as about 1 nm. (b) An atomic resolution image (1.7 V, 50 pA) of a few-layer MoS$_2$ at V$_{\rm g}=40$ V showing triangular lattice with lattice constant close to 0.3 nm. The plot in the inset depicts the topography along the marked line with atomic corrugation.}
	\label{fig:mos22}
\end{figure*}
The STM images were taken in constant current mode and at a $V_{\rm b}$ value well above the conduction band edge of MoS$_2$ and also at a positive $V_{\rm g}$ above the bulk conduction threshold. The latter ensures the tunnel current conduction from the tunnel junction to the Au contacts with an in-between voltage drop much less than the applied $V_{\rm b}$. Further, the junction resistance is typically kept as 20 G$\Omega$ or higher for imaging which is significantly larger than the typical off-state bulk resistance of about 1 G$\Omega$. The local $dI/dV_{\rm b}$ at large negative $V_{\rm g}$ values is found to be still measurable when MoS$_2$ is in the off-state. This is presumably due to the presence of defects and traps that lead to a non-infinite resistance and smaller than tunnel junction resistance. This made it possible to acquire $dI/dV_{\rm b}$ data, as discussed later, in a rare p-doped region. Only a very few \cite{mid-gap states,zhou-STM-TIBB} STM studies of MoS$_2$ on insulating substrates have been reported so far as opposed to those on conducting metals \cite{gap} presumably due to the complication arising from lack of tunnel current conduction in the off state. The off-state of an ultra-clean and trap-free MoS$_2$ can be expected to be highly insulating which can make such STM/S studies even more challenging.

\section{Results and Discussions}
\subsection{ STM topography and atomic resolution image and local $dI/dV_{\rm b}$ spectra }
Figure \ref{fig:mos22}(a) shows a topographic image of a single-layer MoS$_2$ on SiO$_2$/Si substrate acquired using constant current mode with $I=50$ pA, $V_{\rm b} = 2$ V and $V_{\rm g} = +30$ V. This observed corrugation of about 1 nm is same as that of the thermal SiO$_2$ on Si as reported by AFM measurements \cite{AFM}. This could indicate that single layer MoS$_2$ adheres well to the SiO$_2$ surface although some of this corrugation can also be attributed to the electronic inhomogeneities. We also found that few layer MoS$_2$ exhibits a substantially smaller corrugation. A zoomed-in atomic resolution image, see Fig. \ref{fig:mos22}(b), taken on relatively flat region of a few-layer MoS$_2$ shows a triangular lattice of sulfur surface atoms with a lattice constant of about 0.3 nm, in good agreement with the literature \cite{atomic,atomic1}.

\par
\begin{figure*}
	\centering
	\includegraphics[width=15.6cm]{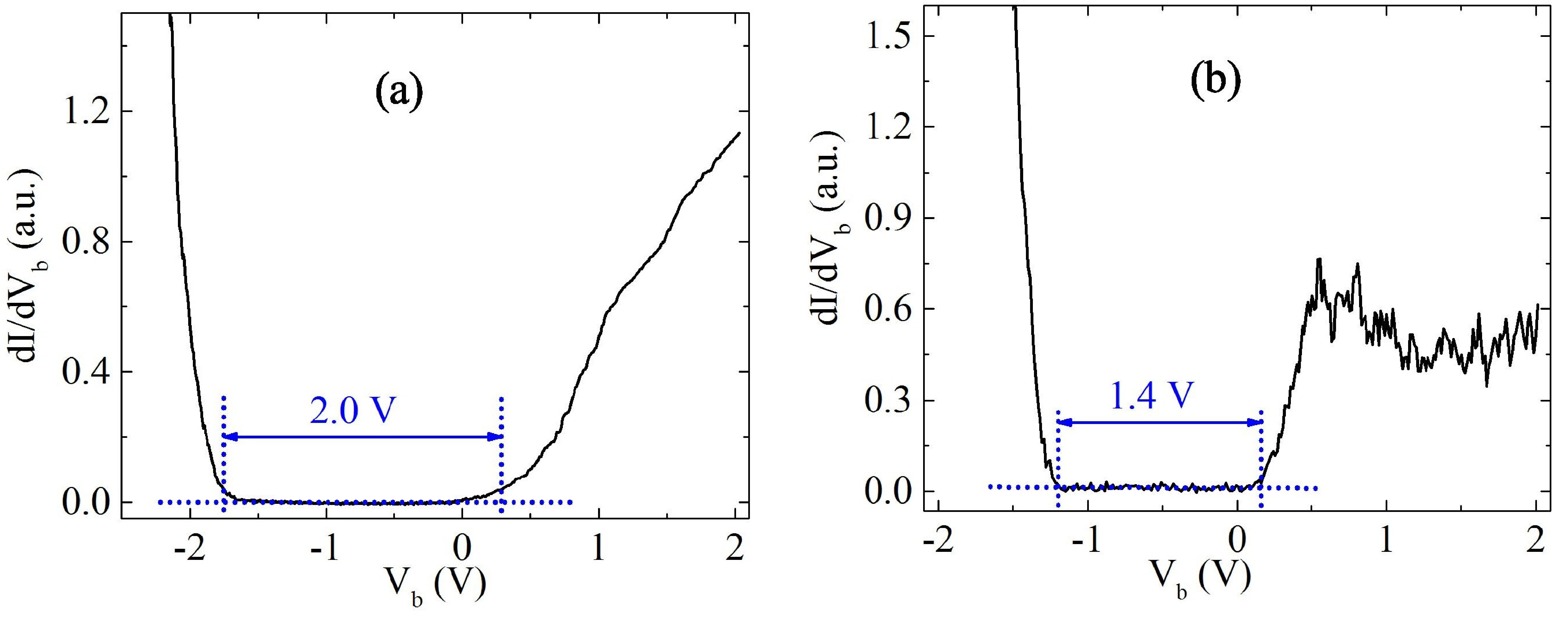}
	\caption{The local tunnel spectra (1.5 V, 100 pA) of (a) Single and (b) few-layer MoS$_2$ acquired at V$_{\rm g}=30$V showing the n-doped nature as deduced from the $V_{\rm b}=0$ point being close to the CBM. As marked, the band gap is approximately $\sim$2.0 eV, and $\sim$1.4 eV for single and few-layer MoS$_2$, respectively.}
	\label{fig:mos23}
\end{figure*}

 Figure \ref{fig:mos23} shows the $dI/dV_{\rm b}-V_{\rm b}$ spectra acquired on the single layer and few-layer MoS$ _{2}$ surfaces at V$_{\rm g} = 30$ V. No hysteresis was found in these spectra with respect to the $V_{\rm b}$-sweep direction. These spectra were acquired several microns away from the metal contacts to avoid their influence. The spectra on both single as well as few-layer MoS$_2$ exhibit a clear gap. For the used bias configuration, see Fig. \ref{fig:mos21}(b), the sharp rise at $V_{\rm b}>0$ in these spectra corresponds to the conduction band minimum (CBM) and the one at $V_{\rm b}<0$ represents the valence band maximum (VBM) with the separation between the two edges representing the band-gap. The marked band-gaps of about 1.4 eV and 2.0 eV for few-layer and single-layer MoS$_2$, respectively, in figure \ref{fig:mos23}, agree with the reported values \cite{gap}. The $V_{\rm b}=0$ point corresponds to the Fermi energy of MoS$_2$, which is close to the CBM in figure \ref{fig:mos23} indicating that both these few-layer and single layer MoS$_2$ flakes are n-doped semiconductors. Electron rich sulfur vacancies and other n-type dopant impurities in natural MoS$_2$ crystals are believed to be responsible for this n-type nature \cite{native defects, channel length,contact}.

The observed tunneling band-gap can actually be significantly larger than the actual one due to the tip-induced band bending (TIBB) \cite{TIBB, TIBB1, TIBB2}, the term used in the framework of 3D semiconductors. Further, the same effects, if significant, would also lead to hysteresis in $dI/dV_{\rm b}-V_{\rm b}$ spectra as is the case with respect to $V_{\rm g}$ which is discussed later. This can be expected as the local band bending will change the occupancy of the slow traps \cite{trap-block-trans} in the vicinity. Lack of such hysteresis and the fact that the observed tunneling band-gap is close to the actual value indicate that effects due to TIBB are insignificant. This can be attributed to the following observations. 1) The occupancy of the traps, arising from defects, changes in response to $V_{\rm b}$ change and amount to screening of tip's electric field that would reduce the band-bending. A significant density of states of fast traps $\sim 10^{12}$ eV$^{-1}$cm$^{-2}$ can be inferred in such devices from the large values of the subthreshold swing \cite{trap-block-trans}. A large density of slow traps can also be deduced from the typical hysteresis in the channel transport in such devices \cite{trap-block-trans}. 2) The back gate is capacitively coupled to the channel with a capacitance equivalent to a quantum capacitance corresponding to a density of states $\kappa\epsilon_0/(e^2d)$ with $\kappa$ as dielectric constant of SiO$_2$ gate, $d$ as its thickness and $\epsilon_0$ as free space permittivity. This works out to be $7.4\times10^{10}$ eV$^{-1}$cm$^{-2}$. This will also contribute towards reducing the TIBB as some of the electric-field lines due to the tip bias will terminate on the gate rather than the channel. 3) The STM tip size is much smaller than the typical screening length within the 2D-MoS$_2$. The latter will dictate the channel area over which the TIBB will spread. This will substantially reduce the local TIBB under the STM tip. In contrast, the back-gate voltage in the 2D-FET configuration affects whole of the 2D MoS$_2$ uniformly leading to significant band-bending. 4) It has been recently argued \cite{zhou-STM-TIBB} that presence of lateral tunneling inside MoS$_2$ will also reduce the TIBB.

In figure \ref{fig:mos23}, the sharpness of the spectra at the VBM is seen to be substantially more than that near CBM for both cases. This can be attributed to the band tail states (BTS) below the CBM. These BTS near CBM can reduce the effective band gap and severely affect the transport. The U-shaped spectra indicating BTS or band edge states near both VBM and CBM have been commonly seen in Si/SiO$_2$ and other conventional 3D semiconductors \cite{U}. The presence of BTS near the CBM can arise from the presence of sulphur vacancies in MoS$_2$ \cite{defect,intrinsic origin,native defects,contact} which are the most common defects. Although a single S vacancy state is expected to be 0.46 eV below the CBM by density functional theory calculations \cite{defect} but we did not find any peaks inside the band gap in our $dI/dV_{\rm b}$ spectra even in the vicinity of defects inferred from the STM images. The observed continuum of BTS close to the CBM could arise from multiple S vacancies that interact with each other and possibly with other defects.

\subsection{Spatial variation of hysteresis in local $dI/dV_{\rm b}$ with gate-sweep.}
\begin{figure*}[h]
	\centering
	\includegraphics[width=15cm]{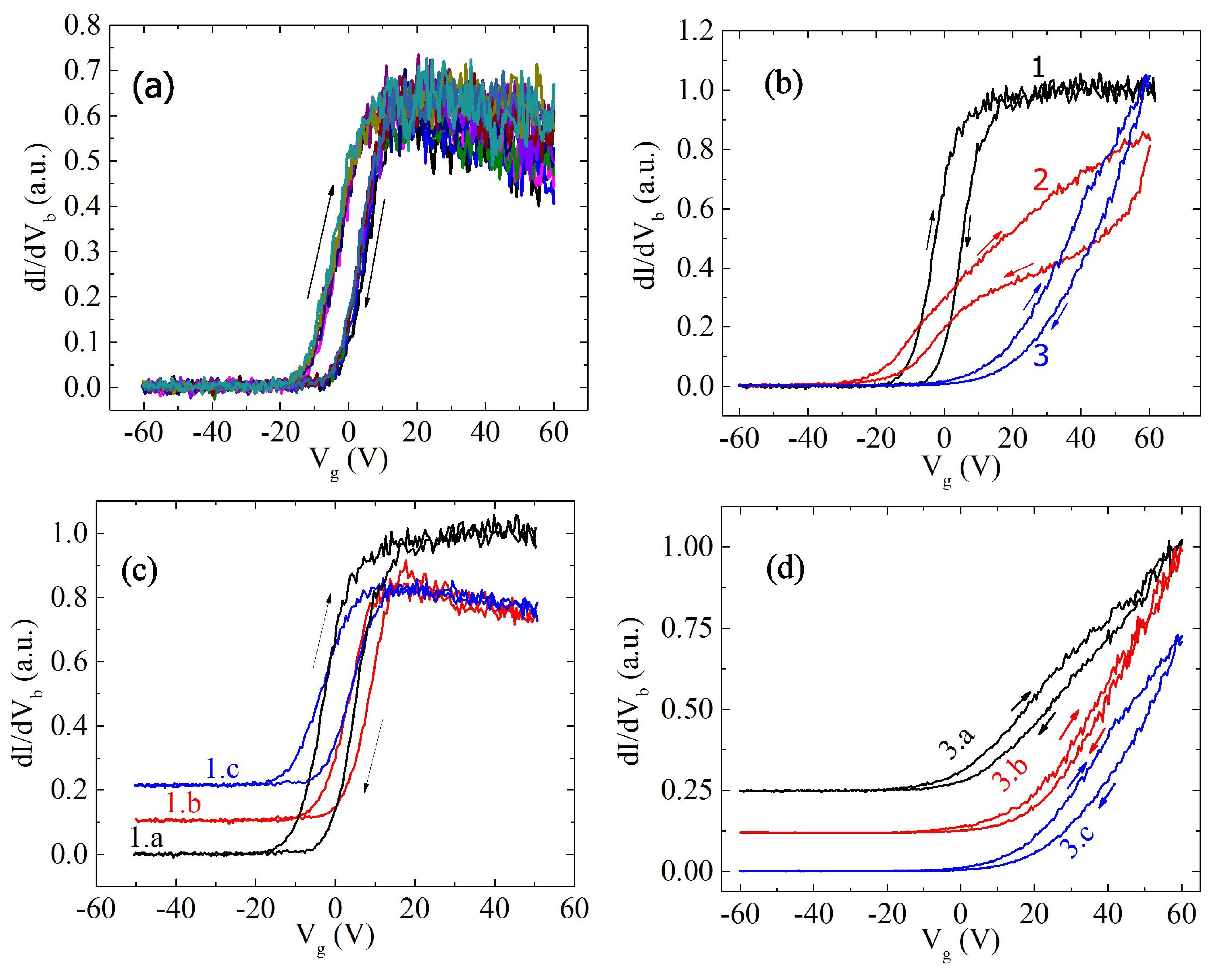}
	\caption{(a) Shows twelve, corresponding to six $V_{\rm g}$ cycles, of unprocessed local $dI/dV_{\rm b}$-$V_{\rm g}$ curves, each exhibiting the same hysteresis, at a fixed $V_{\rm b}=2$ V and 100 pA tunnel current set-point. (b) Shows the average of six cycles of $dI/dV_{\rm b}$-$V_{\rm g}$ curves in three different regions, separated by more than few hundred nm, of single layer MoS$_2$. (c),(d) show $dI/dV_{\rm b}$-$V_{\rm g}$ curves at points that are 20 nm apart and along a straight line in regions `1' and `3' to illustrate variations in $V_{\rm th}$ and $\Delta V_{\rm th}$ within these regions.}
	\label{fig:mos24}
\end{figure*}

\begin{figure*}[h]
	\centering
	\includegraphics[width=15.6cm]{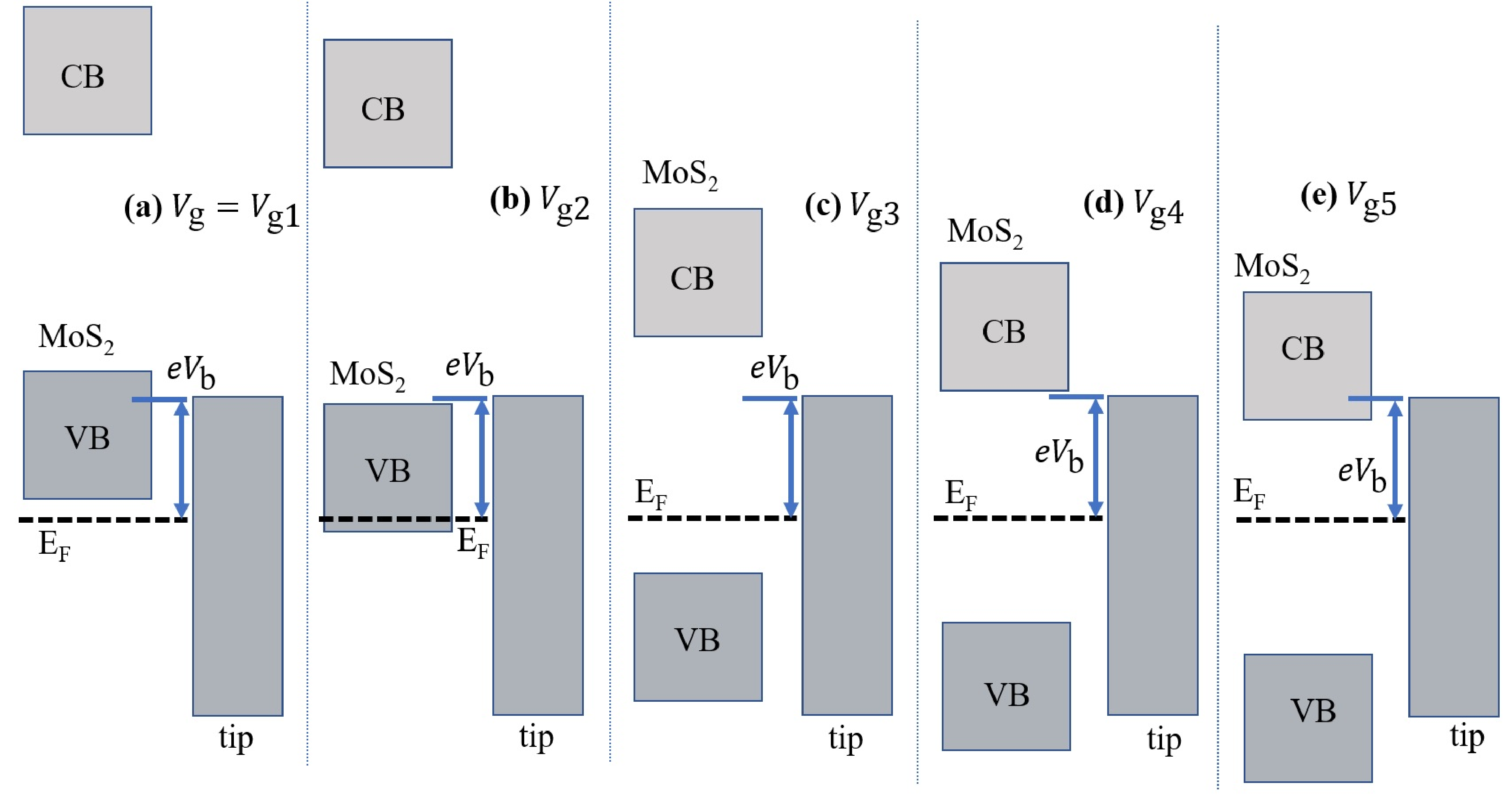}
	\caption{Schematic evolution of MoS$_2$ conduction (CB) and valence band (VB) with increasing $V_{\rm g}$, from (a) to (e), and for a fixed $V_{\rm b}$. Here, $V_{\rm g1}<V_{\rm g2}<V_{\rm g3}<V_{\rm g4}<V_{\rm g5}$ and the traps are assumed to have equilibrium occupancy leading to a monotonic downward shift in bands with increasing $V_{\rm g}$. We use a convention where the MoS$_2$ Fermi energy $E_{\rm F}$ and that of the tip, i.e. $E_{\rm F}+eV_{\rm b}$, remain fixed while the MoS$_2$ bands shift with $V_{\rm g}$ change. In schematic (a) and (e), the tip Fermi energy faces the MoS$_2$ valence and conduction band, respectively, leading to a large $dI/dV_{\rm b}$. (b) represents a threshold when $dI/dV_{\rm b}$ just diminishes with increasing $V_{\rm g}$. Thus, the tip Fermi energy is slightly above, but within a few $k_{\rm B}T$ of, the MoS$_2$ VBM energy leading to the conductance onset due to electrons tunneling from tip to the VB states. (c) represents a zero $dI/dV_{\rm b}$ state while (d) corresponds to a threshold condition where $dI/dV_{\rm b}$ starts rising from zero as tunneling with the CB states begins with increasing $V_{\rm g}$. Please note that at a given sample location all the five scenarios may not be accessible within the experimentally possible $V_{\rm g}$ range due to a large trap density that reduces the shift with $V_{\rm g}$.}
	\label{fig:mos27}
\end{figure*}

Figure \ref{fig:mos24}(a) shows a set of twelve unprocessed individual $dI/dV_{\rm b}$ versus $V_{\rm g}$ curves at a specific location of single layer MoS$_2$ at a fixed $V_{\rm b}=2$ V and for forward and reverse $V_{\rm g}$ sweeps between -60 and 60 V. The schematics in Figure \ref{fig:mos27} illustrate the underlying physics. At $V_{\rm b}= 2$ V, the Fermi level of MoS$_2$ is 2 eV below that of the tip. The Fermi level of the tip, kept at virtual ground, is used as zero energy reference. As $V_{\rm g}$ is increased, at fixed $V_{\rm b}$, from negative to positive during the forward sweep, the bands of MoS$_2$ will shift downward, see fig. \ref{fig:mos27}. At certain threshold $V_{\rm g}$ value, namely $V_{\rm th}$, the tip's Fermi level will be within a few $k_{\rm B}T$ below the MoS$_2$ conduction band edge, see Fig. \ref{fig:mos27}(d), leading to a sudden rise in $dI/dV_{\rm b}$ as the electron start tunneling from the tip to the MoS$_2$ CB states. In Fig. \ref{fig:mos24}, the threshold voltage $V_{\rm thf}$ for the forward $V_{\rm g}$ sweep is seen to be smaller than $V_{\rm thb}$, i.e threshold for the backward sweep. This amounts to a significant positive local hysteresis. Comparing with the bulk transport in similar and those with passivated interface samples \cite{Jana et al}, this observed hysteresis is attributed to the local traps at the interface of MoS$_2$ and SiO$_2$. When $V_{\rm g}$ is increased from negative extreme value, during the forward sweep, the interface traps are positively charged which induces a negative charge on MoS$_2$. This brings MoS$_2$ CBM closer to its Fermi energy as compared to when there is no trap charge and thus a smaller $V_{\rm g}$ is needed to reach the threshold condition. For the backward sweep the traps will be negatively charged, which makes the threshold higher.

A trap-state is characterized by a potential barrier, which determines how fast it exchanges electrons with the channel \cite{Jana et al}. The fast traps, which are strongly coupled to the channel, exchange electrons faster than the experimental time scale. These fast-traps act like dopants as their occupancy gets equilibrated fast and is dictated by the Fermi-energy of MoS$_2$. Further, their local areal density and energy distribution will determine the local $V_{\rm th}$ value of the $dI/dV_{\rm b}$-$V_{\rm g}$ curves but the fast traps will not contribute to the hysteresis. The extremely slow traps, with a large potential barrier, also do not contribute to the hysteresis. The slow traps that exchange electrons with the channel at a time scale comparable to the $V_{\rm g}$ sweep time dominantly contribute to the positive hysteresis.

Assuming only one electron is captured or released per trap, the trap density responsible for the observed hysteresis in figure \ref{fig:mos24}(a) can be estimated as: $N_{\rm tr}= (V_{\rm thb}-V_{\rm thf})C_{\rm ox}/e\approx 7 \times10^{\rm 11}$cm$^{\rm -2}$. Here, $C_{\rm ox} =\kappa\epsilon_{\rm 0}/d\approx$ 12 nF/cm$^{\rm 2}$ is the capacitance of SiO$_2$. The few-layer MoS$_2$ is found to exhibit larger hysteresis and less steep changes in $dI/dV_{\rm b}$-$V_{\rm g}$ curves than mono-layer MoS$_2$ for the same sweep parameters. This is, presumably, due to more defects and inter-layer traps. Note that the interface trap-states do not directly contribute to the tunnel conductance as the direct tunneling matrix element between the tip and interface-trap states will be negligible as compared to that between tip and MoS$_2$ states. The tunneling electrons' equilibration rate is expected to be much faster than the typical electron transfer rate between MoS$_2$ and the trap states. Thus the interface-traps affect the local $dI/dV_{\rm b}$ only through the filling or the band-shift of the MoS$_2$ channel.

Different scales of inhomogeneities were observed in the sense of seeing the same nature of curves over scales of 100$\times$100 nm$^{\rm 2}$ order but with variation in $V_{\rm th}$ values, while over larger scales a change in the nature of these curves is also observed. Figure \ref{fig:mos24}(b) displays three typical $dI/dV_{\rm b}$-$V_{\rm g}$ curves in different regions, more than 100 nm apart. Other than the variation in $V_{\rm th}$ and $\Delta V_{\rm th}=V_{\rm thb}-V_{\rm thf}$, the details of the $dI/dV_{\rm b}$ spectra markedly change for the same $V_{\rm g}$ sweep parameters. These regions do not have any correlation with the topography images. Figures \ref{fig:mos24}(c) and (d) show representative $dI/dV_{\rm b}$-$V_{\rm g}$ curves from regions `1' and `3', respectively, which are taken at points 20 nm apart and along a straight line. Region `1' exhibits a sharper turn-on than Region `3' while region '2' shows a plateau-like feature in backward $V_{\rm g}$ sweep. The $V_{\rm th}$ values of forward sweep in these regions differ, by more than 18 V, which is more than that within a region. This implies a variation in trap density of $\approx $ 1.3$\times10^{\rm 12}$ cm$^{\rm -2}$ over large scales.

Figure \ref{fig:mos27} illustrates how the bands shift relative to the tip Fermi energy when $V_{\rm g}$ is increased. Note that we use a convention where the MoS$_2$ Fermi energy $E_{\rm F}$ and that of the tip, i.e. $E_{\rm F}+eV_{\rm b}$, are kept fixed. The MoS$_2$ bands and the trap-states (not shown in Fig. \ref{fig:mos27}) shift together downward in response to a $V_{\rm g}$ increase. This dictates the Fermi energy of the combined trap and MoS$_2$-band system through the charge induced on this system by back-gate. Thus the trap states effectively screen the gate electric field and slow down the shift of the bands in response to $V_{\rm g}$. The $dI/dV_{\rm b}$ value at a given $V_{\rm g}$ reflects the thermally smeared density states of MoS$_2$ at energy $eV_{\rm b}$ above $E_{\rm F}$. The traps that contribute to hysteresis are the slow ones but both, the fast and slow, will contribute to the shift of the bands and thus to the details of $dI/dV_{\rm b}$-$V_{\rm g}$ curves.

\begin{figure*}
	\centering
	\includegraphics[width=15.6cm]{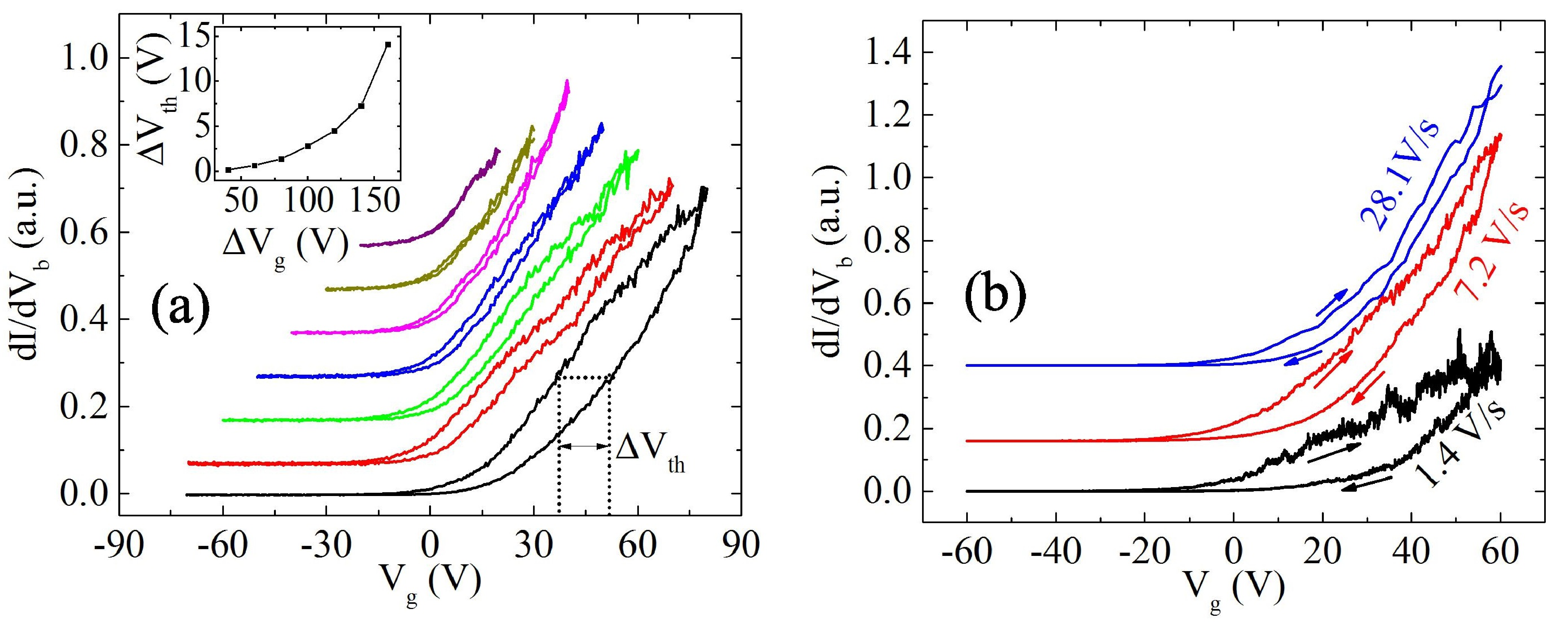}
	\caption{(a) Variation in hysteresis in local dI/d-$V_{\rm g}$ curves with $V_{\rm g}$-sweep range at fixed sweep-rate, bias voltage and tunnel current (2.0 V, 100 pA). Inset shows the variation of $\Delta V_{\rm th}$ with $V_{\rm g}$-sweep range, i.e. $\Delta V_{\rm g}$ as extracted from the plots in (a). (b) Variation in hysteresis in local $dI/dV_{\rm b}$-$V_{\rm g}$ curves with $V_{\rm g}$-sweep rate at fixed sweep range, bias voltage and tunnel current (2.0 V, 100 pA).}
	\label{fig:mos25}
\end{figure*}
In region '1', a sharp rise in $dI/dV_{\rm b}$ at the threshold and near saturation afterwards implies a sharp rise or a peak in traps' density of states. When the Fermi energy of MoS$_2$ and traps is well above this peak the bands and the traps states move down fast, in response to $V_{\rm g}$ increase. This leads to a sharp rise in $dI/dV_{\rm b}$, until the traps' peak reaches $E_{\rm F}$ where the bands stop shifting with $V_{\rm g}$ and $dI/dV_{\rm b}$ stays nearly constant. The $dI/dV_{\rm b}$ can also decrease in case of the delayed reaction of the slow traps which can cause an upward shift with time of the bands with a fixed, or even with an increasing, $V_{\rm g}$. In region '3', on the other hand, the rise in $dI/dV_{\rm b}$ is slow and without any saturation indicating a relatively constant trap density of states leading to a slow downward shift in bands with increasing $V_{\rm g}$. In region '2' the plateau-like feature can arise from a small peak in the trap density of states that checks the shift of bands over a range of $V_{\rm g}$. Furthermore, traps with energies within a few k$_{\rm B}T$ of $E_{\rm F}$ can randomly exchange electrons with the channel leading to the generation-recombination (g-r) noise \cite{g-r noise} in $dI/dV_{\rm b}$, see figure \ref{fig:mos24}(b).

An inhomogeneous distribution of traps can be understood as arising from various non-uniform distribution of defects. These defects include amorphous SiO$_2$ surface defects, surface dangling bonds, surface absorbers, immobile ionic charges, and foreign impurities adsorbed on the surface. Two major SiO$_2$ surface structures have long been recognized: a high-polarization silanol group (Si-OH) and a weak-polarization siloxane group (Si-O-Si) \cite{chemistry, chemistry1}. Negatively charged silanol groups will induce a positive charge on MoS$_2$ and increase local $V_{\rm th}$. H$^{\rm +}$ ions of water molecules hydrogen bond with the OH$^{\rm -}$ of the silanol while exposing O$^{\rm 2-}$ to MoS$_2$, which can exchange electron with MoS$_2$ and alter $V_{\rm th}$. Water molecules on MoS$_2$ cannot be completely removed by pumping in a vacuum at room temperature even for extended periods \cite{chemistry}. The weak polarization siloxane groups (Si-O-Si) lead to hydrophobic nature \cite{chemistry1} as no hydrogen bonding with water molecule is possible. The MoS$_2$ in such regions will exhibit a lower $V_{\rm th}$ as compared to that on the silanol surface.

In addition, foreign impurity atoms present on the SiO$_2$/MoS$_2$ interface can also act as traps. A shallow donor trap state is produced just below the CBM of the MoS$_2$ when a foreign impurity atom is absorbed on the siloxane surface \cite{p-type}. Even at low temperatures and with a low activation energy, it can transfer an electron to the MoS$_2$ conduction band and affect $V_{\rm th}$. In addition, intrinsic defects like sulphur vacancies, as discussed earlier, also affect $V_{\rm th}$. These are more likely to act as fast traps as these will be well coupled to MoS$_2$.

\subsection{Variation of hysteresis with gate voltage sweep range and sweep rate.}
Figure \ref{fig:mos25}(a) shows the $dI/dV_{\rm b}$-$V_{\rm g}$ curves for different $V_{\rm g}$ sweep ranges varying from $\pm20$ V, i.e. $\Delta V_{\rm g}=40$ V to $\pm$80 V, i.e. $\Delta V_g=160$ V. Negligible hysteresis is seen in the $\pm20$ V sweep range and hysteresis, quantified by $\Delta V_{\rm th}=V_{\rm thb}-V_{\rm thf}$, increases non-linearly with $\Delta V_{\rm g}$, see the inset of figure \ref{fig:mos25}(a). This is consistent with the bulk transport in similar devices \cite{Jana et al}. For smaller $\Delta V_{\rm g}$, the MoS$_2$ and traps Fermi energy changes by smaller amount and thus slow-traps in relatively narrower energy range change their charge state as compared to larger $\Delta V_{\rm g}$. We can estimate the number of traps that change their charge state using $\Delta V_{\rm th}$, as discussed earlier. Thus, for $\Delta V_g=160$ V, $1.05\times 10^{\rm 12}$ cm$^{-2}$ traps change their charge state as compared to $6.58\times 10^{\rm 10}$ cm$^{-2}$ at $\Delta V_g=40$ V in figure \ref{fig:mos25}(a).
\begin{figure*}
	\centering
	\includegraphics[width=15.0cm]{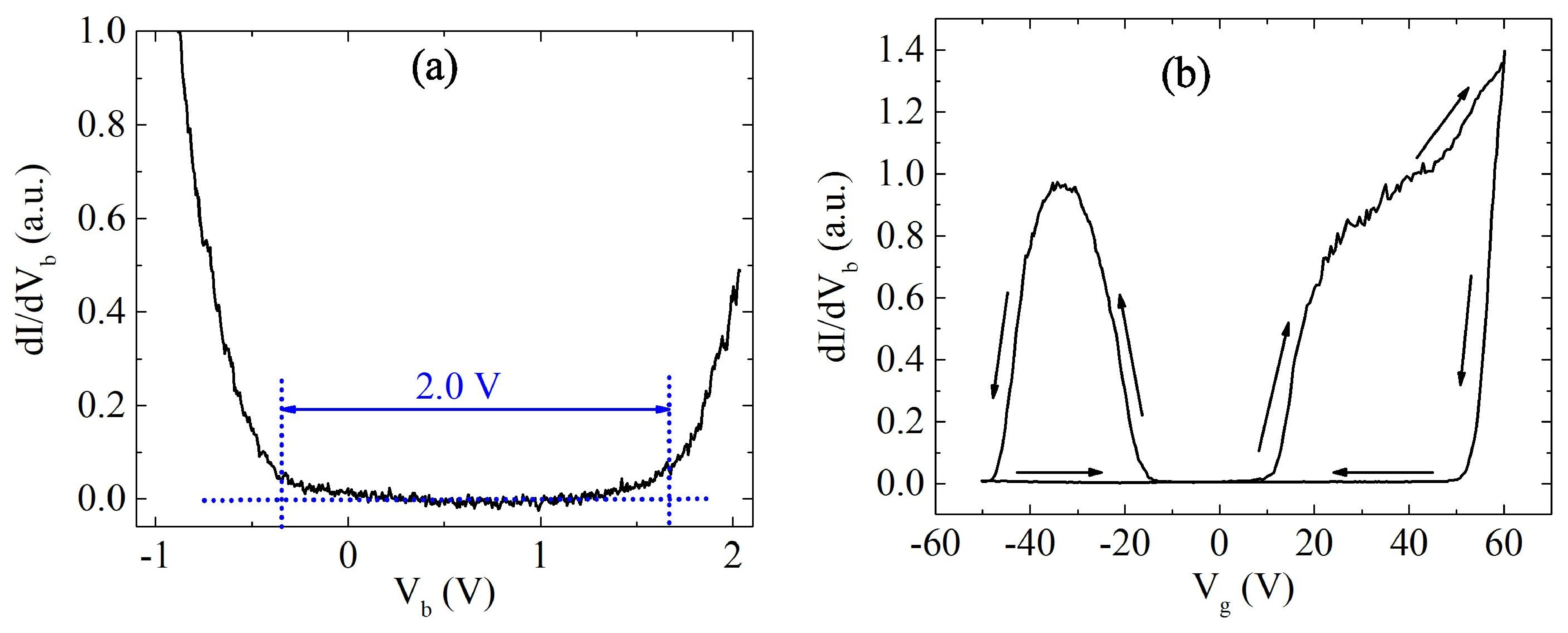}
	\caption{(a) Local tunnel spectra ($V_{\rm b}= 2.0$ V and $I=100$ pA) at certain location of the single layer MoS$_2$ and at V$\rm g=30$ V showing a band-gap of $\sim2eV$ and with Fermi energy close to the valance band edge indicating a p-doping. (b) $dI/dV_{\rm b}$-$V_{\rm g}$ curves (averaged over six sweeps for each direction) at the same location and at a fixed tunnel current and bias voltage (100 pA, 2 V) show an ambipolar nature and with an unusual hysteresis.}
	\label{fig:mos26}
\end{figure*}

\par Figure \ref{fig:mos25}(b) shows the change in hysteresis curves with $V_{\rm g}$ sweep rate for fixed $\Delta V_{\rm g}$ and other parameters. At higher sweep rate more slow-traps will be able to change their charge state. This makes the hysteresis increase with reducing $V_{\rm g}$-sweep rate. A larger sweep-rate also increases the maximum value of the tunnel conductance as reduced traps' participation leads to higher change in channel Fermi energy or carrier density. Also, the slowly acquired $dI/dV_{\rm b}$-$V _{\rm g}$ curves contain more generation-recombination noise, see Fig. \ref{fig:mos25}(b), as more traps capture and release electrons randomly during the slow sweep

\subsection{Observation of a rare hole-doped nature in local tunnel conductance}
Figure \ref{fig:mos26}(a) shows a tunnel spectra, i.e. $dI/dV_{\rm b}-V_{\rm b}$ at certain location with the bias voltage sweep at a fixed $V_{\rm g} = 30$ V. This shows the expected band gap but more band tail states near both VBM and CBM and with MoS$_2$ Fermi energy close to the valance band edge. This indicates a local p-type doping as opposed to the n-type which is more common. If the underlying SiO$_2$ is defect-free, it has also no influence on MoS$_2$'s local conductance since its valence and conduction bands are far apart from those of MoS$_2$ \cite{p-type}. No charge transfer occurs between SiO$_2$ and MoS$_2$ due to the high hole and electron barrier. Hence, the observed n-or p-type local conductance of MoS$_2$ on SiO$_2$ must arise from defects, impurities and local disorder at the MoS$_2$-SiO$_2$ interface. Regions with sulfur-rich or Molybdenum-deficiency can also make it p-type locally. The traps may also give rise to this behavior depending on their energy relative to the MoS$_2$ bands. Thermally created undercoordinated oxygen atoms (Si-O*) on the silanol (Si-OH) terminated surface can serve as an electron trap leading to p-doping in MoS$_2$. For such a defect it is reported that the empty acceptor state is generated at 0.9 eV above the SiO$_2$ VBM \cite{p-type}.

\par Figure \ref{fig:mos26}(b) shows $dI/dV_{\rm b}$-$V_{\rm g}$ curves at the same location with a rather unusual hysteresis and with access to both electron and hole doped regimes through $V_{\rm g}$ change. During the forward sweep, local conductance becomes non-zero at $V_{\rm g} = 10$ V and increases till $V_{\rm g}=60$ V and then it quickly drops to zero at the beginning of the backward sweep followed by a p-type threshold at around $V_{\rm g} =-10$ V. On further decreasing $V_{\rm g}$, conductance increases till $V_{\rm g} = -35$ V and then decrease again reaching zero at -50 V which stays at zero at the begining of forward sweep of $V_{\rm g}$. The nono-monotonic change in $dI/dV_{\rm b}$, and particularly a maximum near $V_{\rm g}=-32$ V can arise from the delayed reaction of the slow traps with energy near VBM which depletes the hole-type carriers from the channel.

Thus in this region of MoS$_2$, all five scenarios of Fig. \ref{fig:mos27} seem to be accessible, presumably, due to a lower overall trap density leading to a wide variation of $E_{\rm F}$ with $V_{\rm g}$ spanning both CBM and VBM. This is more likely from much lower fast-trap density as the hysteresis, arising due to the slow traps, is still quite significant. Also the observation of a finite $dI/dV_{\rm b}$ at a negative $V_{\rm g}$ implies that the bulk of the MoS$_2$ in its insulating state is not too insulating to forbid the transport of tunnel current.
\begin{figure}
	\centering
	\includegraphics[width=8.0cm]{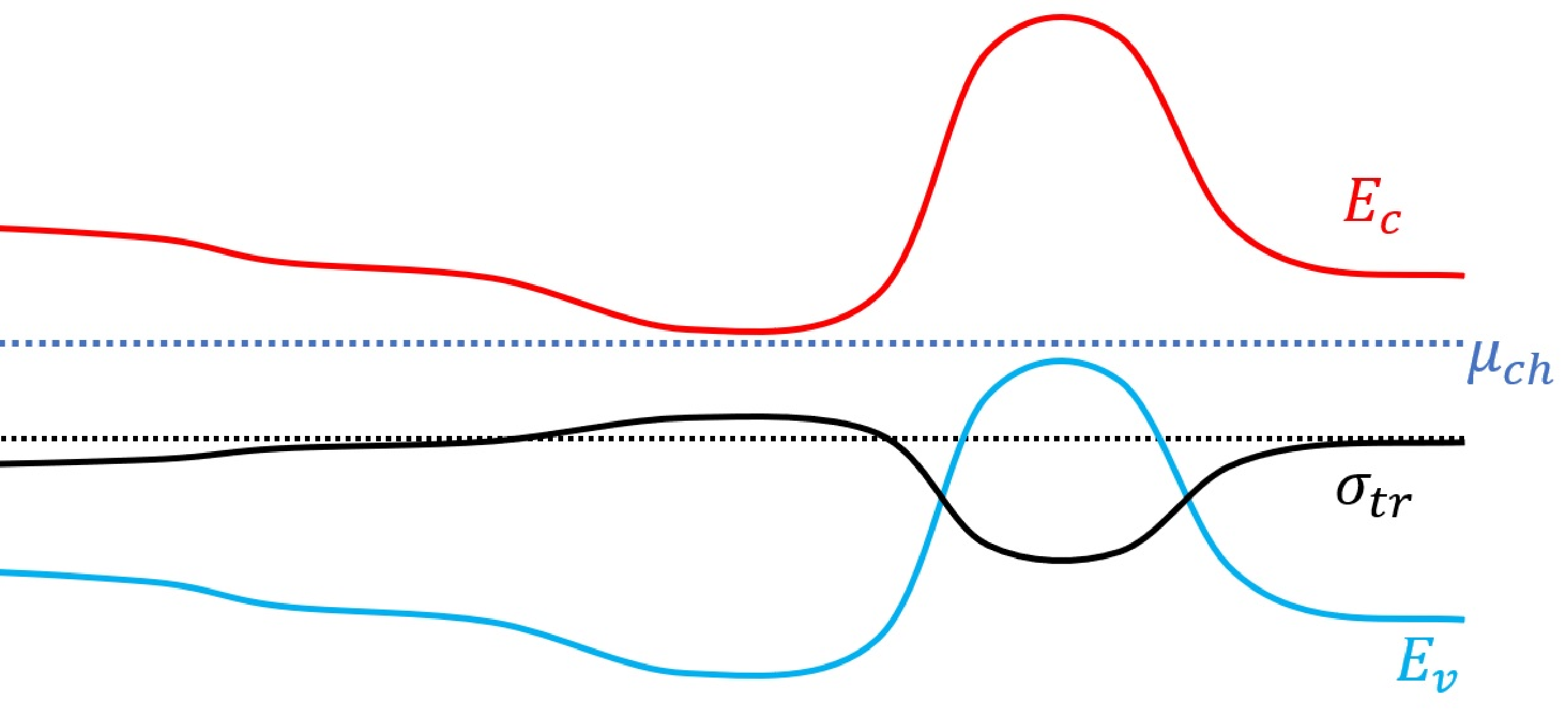}
	\caption{One dimensional schematic of inhomogeneity in trap charge density $\sigma_{\rm tr}$ (black continuous line with black discontinuous one as zero reference line) at equilibrium and the corresponding local band edges' variation. The red and blue lines correspond to CB minimum $E_{\rm c}$ and VB maximum $E_{\rm v}$, respectively. The discontinuous blue line represents the channel chemical potential $\mu_{\rm ch}$. Note the local electron and hole doped regions from the alignment of bands relative to $\mu_{\rm ch}$ that arise from local positive and negative values of $\sigma_{\rm tr}$, respectively.}
	\label{fig:mos28}
\end{figure}

\section{Summary and conclusions}
Figure \ref{fig:mos27} summarized how the MoS$_2$ bands shift in response to $V_{\rm g}$ at a given location with certain trap charge density. The local traps' charge configuration determines the local charge density $\sigma_{\rm tr}$ arising from traps and thus the local band shift of MoS$_2$. This is illustrated in the schematic in figure \ref{fig:mos28}. The overall electrochemical potential stays constant throughout the MoS$_2$ channel while the local filling or electrostatic potential or Fermi energy is inhomogeneous due to variations in local $\sigma_{\rm tr}$. One can thus have locally n- or p-doped regions as seen in this STM study. These schematics, however, do not capture the detailed behavior of traps in terms of their response time and energy distribution. The latter will affect the hysteresis in local $dI/dV_{\rm b}-V_{\rm g}$ curves as well as the detailed $V_{\rm g}$ dependence.

In conclusion, our STM/S investigation on single-layer MoS$_2$ surface shows hysteresis in local tunnel conductance with gate sweep at a constant sample bias voltage. The $\Delta V_{\rm th}$, $V_{\rm th}$ and the shapes of $dI/dV_{\rm b}-V_{\rm b}$ and $dI/dV_{\rm b}-V_{\rm g}$ curves also show inhomogeneities due to traps. Further, the hysteresis changes with $V_{\rm g}$ sweep range and sweep rate. Energy-dependent interface trap-state density is spatially inhomogeneous and even pins the MoS$_2$ Fermi energy in some places. Finally, the dependence of $dI/dV_{\rm b}-V{\rm b}$ and $dI/dV_{\rm b}-V{\rm b}$ on the traps can be a valuable tool for studying traps and defects in 2D materials.
\section*{Acknowledgments}
The authors acknowledge SERB-DST of the Government of India and IIT Kanpur for financial support.

\end{document}